\documentclass[10pt]{article}    
\usepackage{amsmath}
\usepackage{graphicx}
\usepackage{theorem}

\theorembodyfont{\rmfamily}    
\newcommand{\ket}[1]{|#1\rangle}
\newcommand{\bra}[1]{\langle #1 |}

\newcommand{\startproof}{\noindent\emph{Proof.} }
\providecommand{\qed}{\hfill\rule{1.5ex}{1.5ex}}
\newtheorem{lemma}{Lemma}

\newtheorem{remark}[lemma]{Remark}
\newtheorem{corollary}[lemma]{Corollary}

\newtheorem{definition}[lemma]{Definition}
\newtheorem{theorem}[lemma]{Theorem}
\DeclareMathAlphabet{\bm}{OML}{cmm}{b}{it}

\title{Improvement of stabilizer based entanglement distillation
protocols by encoding operators}
\author{Shun Watanabe \and Ryutaroh Matsumoto \and
Tomohiko Uyematsu \and \\
Department of Communications and Integrated Systems, \\
 Tokyo
Institute of Technology, Tokyo 152-8552, Japan}
\date{November 7, 2005}

\begin{document}
\maketitle
\abstract{
This paper presents a method for enumerating all encoding operators
in the Clifford group for a given stabilizer.
Furthermore, we classify encoding operators into the equivalence classes
such that EDPs (Entanglement Distillation Protocol)
constructed from encoding operators in the same equivalence 
class have the same performance.
By this classification, for a given parameter,
the number of candidates for good EDPs
is significantly reduced. As a result, we find the best EDP among EDPs  
constructed from $[[4,2]]$ stabilizer codes.
This EDP has a better performance than previously known EDPs
over wide range of fidelity.
}


\section{Introduction}

In various methods in quantum communication, we have
to share a maximally entangled state.
Bennett et al.~\cite{bennett96a} proposed the entanglement 
distillation protocol (EDP), which is a scheme for sharing a 
maximally entangled state by spatially separated two parties
with local operations and classical communication.
Classical communication in EDPs can be either one-way or
two-way, and two-way EDPs can distill more entanglement than
one-way EDPs. 

In \cite{gottesman-epp,gottesman03,hamada-epp,matsumoto-epp},
the stabilizer based EDP is proposed, which is 
constructed from the quantum stabilizer
code,
and is generalization of the CSS code based EDP \cite{shor01}. 
By using an $[[n,k]]$ stabilizer code,
we can construct EDPs that distill $k$ Bell states from
$n$ Bell states. 
The recurrence protocol \cite{bennett96b} and
the QPA protocol \cite{deutsch98} are special cases of
stabilizer based EDPs,
which are constructed from $[[2,1]]$ stabilizer codes
\cite[Section 4]{matsumoto-epp}.

By now, we arbitrarily choose one of many encoding operators with a
stabilizer based EDP. However, in construction of EDPs
from quantum stabilizers, choice of encoding operators for
stabilizer codes make large differences in 
performances of constructed EDPs.
Even though there exist infinitely many encoding operators for
a given quantum stabilizer, we cannot implement all encoding
operators efficiently.
The reason is as follows.
Any unitary operator can be approximated by using only 
elementary operators, the Hadamard operator, the phase operator,
the controlled not operator, and the $\frac{\pi}{8}$ operator.
However in general, most unitary operators require exponentially
many elementary operators to be approximated in
high accuracy \cite[Section 4.5]{nielsen-chuang}.

The Clifford group  is the set of unitary operators
generated by the Hadamard operator, the phase
operator, and the controlled not operator.
In particular,
each element in the Clifford group that acts on $n$ qubits
is products of at
most $O(n^2)$ previously described three operators
\cite{gottesman98a, gottesman98b,gottesman-d-thesis}.
Thus, for a given stabilizer, encoding operators
in the Clifford group are efficiently implementable.
It is also known that operators in the Clifford group
can be efficiently simulated on a classical computer 
(Gottesman-Knill Theorem) \cite{gottesman98c}.

There is another method to construct two-way EDPs,
which is the permutation based EDP \cite{dehaene03a}.
Permutation based EDPs utilize
local operations chosen from the Clifford group,
and it is known that choices of local operations
make difference in performances of permutation based EDPs. 
When encoding operators of stabilizer based EDPs are
restricted to operators in the Clifford group,
the classes of stabilizer based EDPs and permutation based EDPs
are equivalent \cite{hostens05b}.
Elements of the Clifford group are described in terms of
symplectic geometry,
which enable us to enumerate all local operations 
for permutation based EDPs \cite{dehaene03b,hostens05a}.

In this paper, we construct a method for enumerating 
all encoding operators in the Clifford group for a
given stabilizer. Furthermore, we classify encoding 
operators into the equivalence classes such that
EDPs constructed from encoding operators in the
same equivalence class have the same performance.
Such a classification has not been considered for either 
the stabilizer based EDP nor the permutation based EDP until now.
By this classification,
for  given parameters,
the number of candidates for good EDPs is significantly reduced.
For example, in the case of EDPs constructed from the $[[4,2]]$ stabilizer 
code, the number of candidates is reduced by
$1/12288$. It took one week to find the best EDP among
EDPs constructed from the $[[4,2]]$ stabilizer code with computer search,
so we need about $200$ years to find the best EDP without our result.

As a result, we find the best EDP over
wide range of fidelity
among EDPs constructed from $[[4,2]]$ stabilizer code.
This EDP has a better performance than previously known
EDPs over wide range of fidelity. 

This paper is organized as follows.
In Section \ref{preliminary},
we review the stabilizer code and the stabilizer based EDP.
In Section \ref{main-result}, we show our main theorems.
In Section \ref{good-edp}, we show the best EDP found by
using our main theorems.


\section{Preliminary}
\label{preliminary}

In this section, we review the stabilizer
code, the encoding operator of the stabilizer code, 
the construction of entanglement distillation
protocols (EDP) from stabilizer codes, and previously known
results about two-way EDPs and the Clifford group. 
To make our argument general,
we use the $p$--dimensional Hilbert space (qudit) instead of
the two--dimensional space (qubit).

\subsection{Stabilizer code}

In this section, we review the non-binary generalization 
\cite{knill96a,rains97} of the
stabilizer code \cite{calderbank97,calderbank98,gottesman96}.

Let ${\cal H}$ be the $p$-dimensional complex linear spaces with 
an orthonormal basis $\{ \ket{0},\ldots, \ket{p-1} \}$,
where $p$ is a prime number.
We define two matrices $X$ and $Z$ by
\[
X\ket{i} = \ket{i+1 \bmod p}, \;
Z\ket{i} = \omega^i \ket{i}
\]
with a complex primitive $p$-th root $\omega$ of $1$.
The matrices $X$ and $Z$ have the following relation
\begin{eqnarray}
\label{relation-xz}
Z X = \omega X Z.
\end{eqnarray}

Let $\mathbf{Z}_p = \{0$, \ldots, $p-1\}$ with addition and 
multiplication taken
modulo $p$, and $\mathbf{Z}_p^n$ be the $n$-dimensional vector
space over $\mathbf{Z}_p$.
For a vector 
$\vec{a} = (a_1,\ldots,a_n|b_1,\ldots,b_n) \in \mathbf{Z}_p^{2n}$,
let
\[
\mathsf{XZ}^n(\vec{a}) = X^{a_1}Z^{b_1} \otimes
\cdots \otimes X^{a_n}Z^{b_n}.
\]
Note that eigenvalues of $X^{a_i}Z^{b_i}$ are 
powers of $\omega$ for $p \ge 3$,
and $\{ \pm 1, \pm \mathbf{i} \}$ for $p=2$,
where $\mathbf{i}$ is the imaginary unit.
For a vector $\vec{c}=(c_1,\ldots,c_n)  \in \mathbf{Z}_p^n$,
we denote
\begin{eqnarray*}
\ket{\vec{c}} = \ket{c_1} \otimes \cdots \otimes \ket{c_n}.
\end{eqnarray*}

\begin{definition}
Let
\begin{eqnarray}
\label{pauli-group}
{\cal P}_n = \left\{
\omega^i \mathsf{XZ}^n(\vec{a}) \mid i \in \mathbf{Z}_p, 
\vec{a} \in \mathbf{Z}_p^{2n}
\right\}
\end{eqnarray}
for $p \ge 3$,
\begin{eqnarray*}
{\cal P}_n = \left\{
\mu \mathsf{XZ}^n(\vec{a}) \mid \mu \in \{ \pm 1, \pm \mathbf{i} \},
\vec{a} \in \mathbf{Z}_p^{2n} \right\}
\end{eqnarray*}
for $p=2$,
and $S$ a commutative subgroup of ${\cal P}_n$.
The group ${\cal P}_n$ is called the Pauli group
and the subgroup $S$ is called a stabilizer. 
\end{definition}

Suppose that $\{\mathsf{XZ}^n(\vec{\xi}_1)$,
\ldots, $\mathsf{XZ}^n(\vec{\xi}_{n-k})$
(and possibly some power of $\omega I_{p^n}$ for $p\ge3$ and
some power of $\mathbf{i} I_{p^n}$ for $p=2$) $\}$
is a generating set of the group $S$,
where $\vec{\xi}_1$, $\ldots$, $\vec{\xi}_{n-k}$ 
are linearly independent over $\mathbf{Z}_p$.
From now, we fix a generating set of $S$ as
$\vec{\xi}_1,\ldots,\vec{\xi}_{n-k}$.

A stabilizer code $Q$ is a joint eigenspace of $S$ in
${\cal H}^{\otimes n}$.
There are many joint eigenspaces of $S$ and
we can distinguish an eigenspace by its eigenvalue
of $\mathsf{XZ}^n(\vec{\xi}_i)$ for $i=1$, \ldots, $n-k$.
Hereafter we fix a joint eigenspace $Q(\vec{0})$ of $S$ and
suppose that $Q(\vec{0})$ belongs to the eigenvalue $\lambda_i$
of $\mathsf{XZ}^n(\vec{\xi}_i)$ for $i=1$, \ldots, $n-k$.
Note that $\lambda_i \in \{ \omega^{a} \mid a \in \mathbf{Z}_p \}$
for $p \ge 3$, and $\lambda_i \in \{ \pm 1, \pm \mathbf{i} \}$
for $p=2$.
For a vector $\vec{x}=(x_1,\ldots,x_{n-k}) \in \mathbf{Z}_p^{n-k}$,
we denote $Q(\vec{x})$ as a joint eigenspace that
belongs to the eigenvalue $\lambda_i \omega^{x_i}$
of  $\mathsf{XZ}^n(\vec{\xi}_i)$ for $i=1$, \ldots, $n-k$.

\begin{definition}
\label{symplectic-inner-product}
For two vectors $\vec{x} =(a_1$,$\ldots$,$a_n|b_1$,\ldots,$b_n)$
and $\vec{y} =(c_1$,\ldots,$c_n|d_1$,\ldots,$d_n)$,
the symplectic inner product is defined by
\begin{eqnarray*}
\langle \vec{x}, \vec{y} \rangle =
\sum_{i=1}^n b_i c_i - a_i d_i.
\end{eqnarray*}
\end{definition}

Suppose that we sent $\ket{\varphi} \in Q(\vec{0})$, and received
$\mathsf{XZ}^n(\vec{e}) \ket{\varphi}$.
We can tell which eigenspace of $S$ contains
the state $\mathsf{XZ}^n(\vec{e}) \ket{\varphi}$ by
measuring
an observable whose eigenspaces are the same
as those of $\mathsf{XZ}^n(\vec{\xi}_i)$.
Then the measurement outcome always indicates
that the measured state $\mathsf{XZ}^n(\vec{e}) \ket{\varphi}$
belonging to the eigenspace  
that belongs to eigenvalue
$\lambda_i \omega^{\langle \vec{\xi}_i,
\vec{e}\rangle}$.

\subsection{Encoding operator}

In this section, we review encoding operators of
stabilizer codes.
An encoding operator of a stabilizer code
is a unitary matrix that maps the canonical basis of 
${\cal H}^{\otimes n}$ to joint eigenvectors
of a stabilizer $S$.

\begin{definition}
Let ${\cal H}^n(\vec{e})$ be the subspace of 
${\cal H}^{\otimes n}$ such that ${\cal H}^n(\vec{e})$ is spanned by
\begin{eqnarray*}
\left\{
\ket{\vec{e}} \otimes \ket{\vec{x}} \mid \vec{x} \in \mathbf{Z}_p^k
\right\},
\end{eqnarray*}
where $\vec{e} = (e_1,\ldots, e_{n-k}) \in  \mathbf{Z}_p^{n-k}$.
\end{definition}

Let $\{ \ket{\vartheta(\vec{e},\vec{x})} \mid \vec{x} \in \mathbf{Z}_p^k \}$
be an orthonormal basis of $Q(\vec{e})$.

\begin{definition}
\label{definition-of-encoding}
An encoding operator $U$ of a stabilizer code is a unitary operator
on ${\cal H}^{\otimes n}$ that maps an orthonormal basis 
of ${\cal H}(\vec{e})$ to an orthonormal basis of $Q(\vec{e})$
for all $\vec{e} \in \mathbf{Z}_p^{n-k}$, i.e.,
\begin{eqnarray*}
U:~{\cal H}(\vec{e}) \ni
\ket{\vec{e}}\otimes \ket{\vec{x}} \mapsto \ket{\vartheta(\vec{e},\vec{x})}
\in Q(\vec{e})
\end{eqnarray*}
for $\vec{e} \in \mathbf{Z}_p^{n-k}$ and 
$\vec{x} \in \mathbf{Z}_p^{k}$.
\end{definition}

Note that
a state $\sum_{\vec{x} \in \mathbf{Z}_p^k} \alpha_{\vec{x}} \ket{\vec{x}}$
of ${\cal H}^{\otimes k}$ is encoded into 
\begin{eqnarray*}
\sum_{\vec{x} \in \mathbf{Z}_p^k} \alpha_{\vec{x}} 
\ket{\vartheta(\vec{e},\vec{x})}
\end{eqnarray*}
by $U$ with ancilla qudits $\ket{e}$.

\subsection{Stabilizer based EDP}

In this section, we review the 
stabilizer based EDP.
We define the maximally entangled states in 
${\cal H}_A^{\otimes n} \otimes {\cal H}_B^{\otimes n}$ by
\begin{eqnarray*}
\ket{\beta^n(\vec{v})} =
I_{p^n} \otimes \mathsf{XZ}^n(\vec{v}) 
\frac{1}{\sqrt{p^n}} \sum_{i=0}^{p^n-1} \ket{i_A} \otimes \ket{i_B},
\end{eqnarray*}
where $\vec{v} \in \mathbf{Z}_p^{2n}$.

Suppose that Alice and Bob share a mixed state 
$\rho \in {\cal S}({\cal H}_A^{\otimes n} \otimes {\cal H}_B^{\otimes n})$,
where 
$\mathcal{S}(\mathcal{H}_A^{\otimes n} \otimes \mathcal{H}_B^{\otimes n})$
is the set of density operators on 
$\mathcal{H}_A^{\otimes n} \otimes \mathcal{H}_B^{\otimes n}$. 
The goal of an entanglement distillation protocol is
to extract as many pairs of particles with state close to 
$\ket{\beta^1(\vec{0})}$
as possible from $n$ pairs of particles in the state $\rho$,
where 
\begin{eqnarray*}
\ket{\beta^1(\vec{0})} = \frac{1}{\sqrt{p}} \sum_{i=0}^{p-1} \ket{i_A}
\otimes \ket{i_B}.
\end{eqnarray*}

For  $\vec{\xi}_i = (a_1,\ldots,a_n|b_1,\ldots,b_n)$,
we define $\vec{\xi}_i^{\star} = (a_1$, $-b_1$, \ldots, $a_n$, $-b_n)$.
Since the complex conjugate of $\omega$ is $\omega^{-1}$,
we can see that $\mathsf{XZ}^n(\vec{\xi}_i^{\star})$ is the
component-wise complex conjugated matrix of $\mathsf{XZ}^n(\vec{\xi}_i)$.
Let $S^\star$ be the subgroup of ${\cal P}_n$ generated by
$\{\mathsf{XZ}^n(\vec{\xi}_1^\star)$,
\ldots, $\mathsf{XZ}^n(\vec{\xi}_{n-k}^\star)\}$.
Easy computation shows that $S^\star$ is again commutative.
So we can consider joint eigenspaces of $S^\star$.
There exists a joint eigenspace $Q^\star(\vec{x})$ of $S^\star$
whose eigenvalue of $\mathsf{XZ}^n(\vec{\xi}_i^{\star})$ is
$\bar{\lambda}_i \omega^{-x_i}$,
where $\bar{\lambda}_i$ is the complex conjugate of $\lambda_i$. 
For a state
\[
\ket{\varphi} = \alpha_0 \ket{0} + \cdots
+\alpha_{p^n-1} \ket{p^n-1} \in \mathcal{H}^{\otimes n},
\]
we define
\[
\overline{\ket{\varphi}} = \bar{\alpha}_0 \ket{0} + \cdots
+\bar{\alpha}_{p^n-1} \ket{p^n-1},
\]
where $\bar{\alpha}_i$ is the complex conjugate of $\alpha_i$.

With those notation,
our protocol is executed as follows.
\begin{enumerate}
\item\label{step1} Alice measures an observable corresponding
to $\mathsf{XZ}^n(\vec{\xi}_{i}^\star)$ for each $i$,
and let $\bar{\lambda}_i \omega^{-a_i}$ be the eigenvalue
of an eigenspace of $S^\star$ containing the state
after measurement.
In what follows we refer to $(a_1$, \ldots, $a_{n-k}) \in \mathbf{Z}_p^{n-k}$
as a \emph{measurement outcome}.

\item\label{step2} Bob measures an observable corresponding
to $\mathsf{XZ}^n(\vec{\xi}_{i})$ for each $i$,
and let $\lambda_i \omega^{b_i}$ be the eigenvalue
of an eigenspace of $S$ containing the state
after measurement.
In what follows we also refer to 
$(b_1$, \ldots, $b_{n-k}) \in \mathbf{Z}_p^{n-k}$
as a \emph{measurement outcome}.

\item Alice sends $(a_1$, \ldots, $a_{n-k})$ to Bob.
\item \label{step-abort}
If the difference of measurement outcomes
$(b_1-a_1,\ldots,b_{n-k}-a_{n-k}) \notin T$ for
a previously specified set $T \subset \mathbf{Z}_p^{n-k}$,
then they abort the protocol.
\item\label{step4} Bob performs the error correction process 
according to $a_1$, \ldots, $a_{n-k}$ as follows:
Bob finds a matrix $M \in {\cal P}_n$ such that $MQ(\vec{b}) = Q(\vec{a})$.
There are many matrices $M$ with $MQ(\vec{b}) = Q(\vec{a})$,
and Bob choose $M$ providing the highest fidelity among those matrices.
See \cite{matsumoto-epp} for details.
He applies $M$ to his particles.
\item\label{step5} Alice and Bob apply the inverse of encoding operators
$\overline{U}^*$ and $U^*$
of the quantum stabilizer codes respectively,
where $U^*$ is the adjoint operator of the encoding operator $U$
and $\overline{U}^*$ is the component-wise complex conjugate operator
of $U^*$.
We stress that Alice applies $\overline{U}^*$ instead of $U^*$
\cite{hamada-epp,matsumoto-epp}.
\item\label{step6} Alice and Bob discards $n-k$ ancilla qudits.
\end{enumerate}

Note that, when we start with the state $\ket{\beta^n(\vec{u})}$,
the state becomes
\begin{equation}
(I_{p^n}\otimes \mathsf{XZ}^n(\vec{u}))
\sum_{\vec{x} \in \mathbf{Z}_p^k} \overline{\ket{\vartheta(\vec{a},\vec{x})}}
\otimes \ket{\vartheta(\vec{a},\vec{x})}
\label{stateafterstep1}
\end{equation}
after Step \ref{step1} \cite[proof of Lemma 1]{matsumoto-epp}.

\subsection{Clifford group}

\begin{definition}
Let ${\cal U}_n$ be the set of all unitary operators
on ${\cal H}^{\otimes n}$, and $N({\cal P}_n)$ be
the normalizer of ${\cal P}_n$ in ${\cal U}_n$, i.e.,
\begin{eqnarray*}
N({\cal P}_n) = \left\{
U \mid U \in {\cal U}_n,~ UMU^* \in {\cal P}_n~ \forall M \in {\cal P}_n
\right\} ,
\end{eqnarray*}
which is called the Clifford group, where $U^*$ is the adjoint operator 
of $U$.
\end{definition}

The unitary operators in the Clifford group $N({\cal P}_n)$
are decomposed into products of the elementary operators,
where elementary operators for $p=2$ are
the Hadamard operator, the phase operator,
and the controlled not operator \cite{gottesman98a,gottesman-d-thesis},
and the elementary operators for $p > 2$ are
the $p$--dimensional discrete Fourier transform operator,
the sum operator, the $p$--dimensional phase operator,
and the $S$ operator \cite{gottesman98b}.
The required number of the elementary operators to represent an operator
in the Clifford group
is at most $O(n^2)$.


\section{Construction of encoding operators}
\label{main-result}

In this section, we present a method to enumerate all  
encoding operators in 
the Clifford group for a given stabilizer 
(Definitions \ref{def-construction} and \ref{enumeration}).
Then, we show relations between Bell states and
encoded Bell states (Lemma \ref{lemma-encoded-Bell},
Corollary \ref{corollary-encoded-Bell}, and Corollary \ref{corollary-inverse}).
Then, we classify encoding operators into equivalence classes such
that EDPs constructed from
encoding operators in the same equivalence class have 
the same performances (Definition \ref{definition_similar},
Theorems \ref{performance-equal} and \ref{performance-equal-2}).
Finally, we show the method to enumerate all equivalence classes
(Theorem \ref{theorem-1}).

\subsection{Construction method}

For a given stabilizer $S$, we define ${\cal M}(S)$ as the
set of all encoding operators, which maps the subspace 
${\cal H}(\vec{e})$ to the subspace $Q(\vec{e})$
for all $\vec{e} \in \mathbf{Z}_p^{n-k}$
(See Definition \ref{definition-of-encoding}).
\begin{definition}
Let ${\cal M}_{\mathrm{cl}}(S)$ be the subset of ${\cal M}(S)$
defined by
\begin{eqnarray*}
{\cal M}_{\mathrm{cl}}(S) = {\cal M}(S) \cap N({\cal P}_n).
\end{eqnarray*}
\end{definition}
${\cal M}_{\mathrm{cl}}(S)$ is the set of all encoding 
operators that are contained in the Clifford group.
The goal of this section is 
to present a method for enumerating all elements of 
${\cal M}_{\mathrm{cl}}(S)$.
Although the method for enumerating all elements of
the Clifford group is known \cite{dehaene03b,hostens05a},
the method for enumerating all elements of ${\cal M}_{\mathrm{cl}}(S)$ for
a given stabilizer $S$ is not known. 

\begin{definition}
The linear space $\mathbf{Z}_p^{2n}$ with symplectic inner product
defined in Definition \ref{symplectic-inner-product} is
called the symplectic space.
\end{definition}

\begin{definition}
Let $\{ \vec{x}_1,\ldots,\vec{x}_n,$
$\vec{y}_1,\ldots,\vec{y}_n\}$ be a basis of 
a  symplectic space $\mathbf{Z}_p^{2n}$.
If $\vec{x}_i $ and $\vec{y}_i$ satisfy
\begin{eqnarray*}
\langle \vec{x}_i, \vec{y}_j \rangle &=& \delta_{ij}, \\
\langle \vec{x}_i, \vec{x}_j \rangle &=& 0, \\
\langle \vec{y}_i, \vec{y}_j \rangle &=& 0
\end{eqnarray*}
for all $i$ and  $j$,
then the basis $\{\vec{x}_1,\ldots,\vec{x}_n,$
$\vec{y}_1,\ldots,\vec{y}_n\}$ is called a hyperbolic basis.
\end{definition} 

\begin{lemma}
There exists vectors $\vec{\xi}_{n-k+1},\ldots,\vec{\xi}_n$
and $\vec{\eta}_1,\ldots,\vec{\eta}_n$ such that
\begin{eqnarray}
\langle \vec{\xi}_i, \vec{\eta}_j \rangle &=& \delta_{ij}, \nonumber \\
\langle \vec{\xi}_i, \vec{\xi}_j \rangle &=& 0, \label{hyperboric}\\
\langle \vec{\eta}_i, \vec{\eta}_j \rangle &=& 0, \nonumber
\end{eqnarray}
i.e., $\{\vec{\xi}_1,\ldots,\vec{\xi}_n,$
$\vec{\eta}_1,\ldots,\vec{\eta}_n\}$ form a hyperbolic basis of
$\mathbf{Z}_p^{2n}$.
\end{lemma}
\startproof
The assertion of this lemma follows from the standard fact 
in symplectic geometry \cite{symplectic-groups, geometric-algebra}. \qed

\begin{lemma}
Let $C$ be a linear subspace of $\mathbf{Z}_p^{2n}$
spanned by $\vec{\xi}_1$, \ldots, $\vec{\xi}_{n-k}$,
and $C^\perp$ be the orthogonal space of $C$
with respect to the symplectic inner product.
Let $C_{\max}$ be a linear subspace of $\mathbf{Z}_p^{2n}$ spanned by
$\vec{\xi}_1,\ldots,\vec{\xi}_n$.
Then,
\begin{eqnarray*}
C_\mathrm{max} &=& C_\mathrm{max}^\perp,\\
C \subseteq & C_\mathrm{max}& \subseteq C^\perp,
\end{eqnarray*}
and $C^\bot$ is spanned by 
$\vec{\xi}_1,\ldots,\vec{\xi}_n, \vec{\eta}_{n-k+1},\ldots,\vec{\eta}_n$.
\end{lemma}
\startproof The assertion of this lemma follows
from the property of a hyperbolic basis. \qed

\begin{definition}
For $p=2$,
we define $\mu(\vec{\xi}_i), \mu(\vec{\eta}_i) \in \{ \pm 1,\pm \mathbf{i} \}$ for 
each $\mathsf{XZ}^n(\vec{\xi}_i), \mathsf{XZ}^n(\vec{\eta}_i)$ as follows,
where $\mathbf{i}$ is the imaginary unit.
For a vector $\vec{\xi}_i = (a_1,\ldots,a_n|b_1,\ldots,b_n)$,
we define $m(\vec{\xi}_i) = |\{i \mid a_i = b_i = 1 \}|$, i.e.,
the number of $XZ$s in $\mathsf{XZ}^n(\vec{\xi}_i)$.
We define $\mu(\vec{\xi}_i)$  as
\begin{eqnarray*}
\mu(\vec{\xi}_j) = \mathbf{i}^{m(\vec{\xi}_i)}.
\end{eqnarray*}
$\mu(\vec{\eta}_i)$ is defined in the same way.
\end{definition}
For example, in case of $n=4$ and
$\mathsf{XZ}^4(\vec{\xi}_j) = X \otimes XZ \otimes XZ \otimes XZ$,
$\mu(\vec{\xi}_j) = -\mathbf{i}$. In case of $n=3$ and 
$\mathsf{XZ}^3(\vec{\xi}_j) = XZ \otimes I_2 \otimes XZ$,
$\mu(\vec{\xi}_j) = -1$.
We need $\mu(\vec{\xi}_j)$ so that
$\left( \mu(\vec{\xi}_j) \mathsf{XZ}^n(\vec{\xi}_j) \right)^2 = I_{2^n}$.
For $p \ge 3$, we do not need $\mu(\vec{\xi}_j)$ and $\mu(\vec{\eta}_j)$.

\begin{definition}
\label{def-construction}
Let $S_\mathrm{max}$ be a subgroup of ${\cal P}_n$
generated by $\{ \mathsf{XZ}^n(\vec{x}) \mid
\vec{x} \in C_\mathrm{max}\}$.
Let $Q_\mathrm{min}(\vec{0})$ be a stabilizer code
defined by $S_\mathrm{max}$ contained in $Q(\vec{0})$.
We have $\dim Q_\mathrm{min}(\vec{0}) = 1$.
Let $\ket{\psi(\vec{0})} \in Q_\mathrm{min}(\vec{0})$
be a state vector of unit norm.

Let 
\begin{eqnarray}
\widetilde{\mathsf{X}}^n(\vec{f}_i) &=&
\theta_x(\vec{f}_i) \mathsf{XZ}^n(\vec{\eta}_i), \\
\widetilde{\mathsf{Z}}^n(\vec{f}_i) &=&
\theta_z(\vec{f}_i) \mathsf{XZ}^n(\vec{\xi}_i)
\end{eqnarray}
for $p \ge 3$, and 
\begin{eqnarray}
\widetilde{\mathsf{X}}^n(\vec{f}_i) &=& \theta_x(\vec{f}_i) 
\mu(\vec{\eta}_i) \mathsf{XZ}^n(\vec{\eta}_i), 
\label{encoded-x-2}\\
\widetilde{\mathsf{Z}}^n(\vec{f}_i) &=& \theta_z(\vec{f}_i) 
\mu(\vec{\xi}_i)\mathsf{XZ}^n(\vec{\xi}_i)
\label{encoded-z-2}
\end{eqnarray}  
for $p=2$, where $\vec{f}_i$ is a vector such that the 
$i$-th element is $1$ and the other elements are $0$,
$\theta_x(\cdot)$ is an arbitrary power of $\omega$,
and we choose $\theta_z(\cdot)$ so that 
$\widetilde{\mathsf{Z}}^n(\vec{f}_i) \ket{\psi(\vec{0})} = 
\ket{\psi(\vec{0})}~~\mbox{for } i = 1,\ldots, n$.

Let 
\begin{eqnarray}
\widetilde{\mathsf{X}}^n(\vec{u}) &=&  
\prod_{i=1}^n (\widetilde{\mathsf{X}}(\vec{f}_i))^{u_i} \label{encoded-x}\\
\widetilde{\mathsf{Z}}^n(\vec{v}) &=&  
\prod_{i=1}^n (\widetilde{\mathsf{Z}}(\vec{f}_i))^{v_i}
\label{encoded-z}
\end{eqnarray}  
for $\vec{u} = (u_1,\ldots,u_n) \in \mathbf{Z}_p^n$ and
$\vec{v} = (v_1,\ldots,v_n) \in \mathbf{Z}_p^n$.

We define our encoding operator $U_e$ by
\begin{eqnarray}
\label{def-encoding-operator}
U_e:~ \mathsf{X}^n(\vec{u})\ket{\vec{0}} \mapsto \widetilde{\mathsf{X}}^n(\vec{u}) \ket{\psi(\vec{0})},
\end{eqnarray}
where $\mathsf{X}^n(\vec{u}) = X^{u_1}\otimes\cdots\otimes X^{u_n}$.
We define $\mathsf{Z}^n(\vec{u})$ in a similar manner.
\end{definition}

\begin{remark}
From Lemma \ref{encoded-x-z}, we find that
Eqs.~(\ref{encoded-x}) and (\ref{encoded-z}) are a generalization of
encoded $\mathsf{X}^n(\vec{u})$ operator and 
encoded $\mathsf{Z}^n(\vec{v})$ operator defined in 
\cite{gottesman-d-thesis}.
\end{remark}

\begin{remark}
The construction of the encoding operator depends
on the choice of $\vec{\xi}_{n-k+1},\ldots,\vec{\xi}_n$
and $\vec{\eta}_1,\ldots,\vec{\eta}_n$
that satisfy Eq.~(\ref{hyperboric}),
$Q_{\min}(\vec{0}) \subset Q(\vec{0})$,
and phase factors $\theta_x(\cdot)$.
An example will be given in  Section \ref{good-edp}.
\end{remark}

\begin{definition}
\label{enumeration}
For a given stabilizer $S$, we define ${\cal M}_g(S)$ as the set of 
encoding operators $U_e$ for 
all choices of $\vec{\xi}_{n-k+1},\ldots,\vec{\xi}_n$,
$\vec{\eta}_1,\ldots,\vec{\eta}_n$,
$Q_{\min}(\vec{0}) \subset Q(\vec{0})$,
and $\theta_x(\cdot)$.
\end{definition}
${\cal M}_g(S)$ is the set of all encoding operators 
that are constructed by the method in Definition \ref{def-construction}.
Next, we show ${\cal M}_g(S)$ is equal to ${\cal M}_{\mathrm{cl}}(S)$.

\begin{lemma}
\label{encoded-x-z}
For $\widetilde{\mathsf{X}}^n(\vec{s}), \widetilde{\mathsf{Z}}^n(\vec{t}),U_e$ defined by Eqs.~(\ref{encoded-x}), (\ref{encoded-z}), 
and (\ref{def-encoding-operator}), we have
\begin{eqnarray}
\label{eq-uxu}
U_e \mathsf{X}^n(\vec{s}) U_e^* &=& \widetilde{\mathsf{X}}^n(\vec{s}) 
\in {\cal P}_n  \hspace{5mm} \forall \vec{s} \in \mathbf{Z}_p^n, \\
\label{eq-uzu}
U_e \mathsf{Z}^n(\vec{t}) U_e^* &=& \widetilde{\mathsf{Z}}^n(\vec{t})
\in {\cal P}_n  \hspace{5mm} \forall \vec{t} \in \mathbf{Z}_p^n.
\end{eqnarray}
\end{lemma}
\startproof
For $\vec{u} \in \mathbf{Z}_p^n$, let 
$\ket{\varphi(\vec{u})}= U_e \ket{\vec{u}} =
\widetilde{\mathsf{X}}^n(\vec{u})\ket{\psi(\vec{0})}$. 
For $\vec{s} \in \mathbf{Z}_p^n$, we have
\begin{eqnarray*}
U_e \mathsf{X}(\vec{s}) U_e^* \ket{\varphi(\vec{u})} &=&
U_e \mathsf{X}^n(\vec{s}) \ket{\vec{u}} \\
&=& U_e \ket{\vec{u}+\vec{s}} \\
&=& \widetilde{\mathsf{X}}^n(\vec{u}+\vec{s}) \ket{\psi(\vec{0})} \\
&=& \widetilde{\mathsf{X}}^n(\vec{s})\widetilde{\mathsf{X}}^n(\vec{u}) 
\ket{\psi(\vec{0})} \\
&=& \widetilde{\mathsf{X}}^n(\vec{s})\ket{\varphi(\vec{u})}.
\end{eqnarray*}
Since $\{ \ket{\varphi(\vec{u})} \mid \vec{u} \in \mathbf{Z}_p^n \}$
form an orthonormal basis of ${\cal H}^{\otimes n}$, we have
\begin{eqnarray*}
U_e \mathsf{X}^n(\vec{s}) U_e^* = \widetilde{\mathsf{X}}^n(\vec{s}) 
\in {\cal P}_n  \hspace{5mm} \forall \vec{s} \in \mathbf{Z}_p^n.
\end{eqnarray*}

Next, for $\vec{t} \in \mathsf{Z}_p^n$, we have
\begin{eqnarray}
U_e \mathsf{Z}^n(\vec{t}) U_e^* \ket{\varphi(\vec{u})} &=&
U_e \mathsf{Z}^n(\vec{t}) \ket{\vec{u}} \nonumber \\
&=& \omega^{(\vec{t},\vec{u})} U_e \ket{\vec{u}} \nonumber \\
&=& \omega^{(\vec{t},\vec{u})} 
\widetilde{\mathsf{X}}^n(\vec{u}) \ket{\psi(\vec{0})},
\label{eq-1}
\end{eqnarray}
where $(\cdot,\cdot)$ is the standard inner product.
From the definition of $\widetilde{\mathsf{Z}}^n(\vec{t})$,
we have $\widetilde{\mathsf{Z}}^n(\vec{t})
\ket{\psi(\vec{0})} = \ket{\psi(\vec{0})}$. 
Since $\vec{\xi}_i$ and $\vec{\eta}_j$
satisfy Eq.~(\ref{hyperboric}), we have 
\begin{eqnarray*}
\mathsf{XZ}^n(\vec{\eta}_i) \mathsf{XZ}^n(\vec{\xi}_j) &=&
\omega^{-1} \mathsf{XZ}^n(\vec{\xi}_j) \mathsf{XZ}^n(\vec{\eta}_i), \\
\mathsf{XZ}^n(\vec{\xi}_i)\mathsf{XZ}^n(\vec{\xi}_j) &=&
\mathsf{XZ}^n(\vec{\xi}_j)\mathsf{XZ}^n(\vec{\xi}_i), \\
\mathsf{XZ}^n(\vec{\eta}_i) \mathsf{XZ}^n(\vec{\eta}_j)
&=& \mathsf{XZ}^n(\vec{\eta}_j)\mathsf{XZ}^n(\vec{\eta}_i),
\end{eqnarray*}
and
\begin{eqnarray*}
\widetilde{\mathsf{X}}^n(\vec{u}) \widetilde{\mathsf{Z}}^n(\vec{t}) 
= \omega^{-(\vec{t},\vec{u})} \widetilde{\mathsf{Z}}^n(\vec{t})
\widetilde{\mathsf{X}}^n(\vec{u}).
\end{eqnarray*}
Since
$\widetilde{\mathsf{Z}}(\vec{t}) \ket{\psi(\vec{0})} =
\ket{\psi(\vec{0})}$,
Eq.~(\ref{eq-1}) is equal to
\begin{eqnarray*}
&& \omega^{(\vec{t},\vec{u})} 
\widetilde{\mathsf{X}}^n(\vec{u}) \widetilde{\mathsf{Z}}^n(\vec{t})
\ket{\psi(\vec{0})} \\
&=& \omega^{(\vec{t},\vec{u})}
\omega^{-(\vec{t},\vec{u})}
\widetilde{\mathsf{Z}}^n(\vec{t}) \widetilde{\mathsf{X}}^n(\vec{u})
\ket{\psi(\vec{0})} \\
&=& \widetilde{\mathsf{Z}}^n(\vec{t}) \ket{\varphi(\vec{u})}.
\end{eqnarray*}
Thus, we have
\begin{eqnarray*}
U_e \mathsf{Z}^n(\vec{t}) U_e^* = \widetilde{\mathsf{Z}}^n(\vec{t})
\in {\cal P}_n  \hspace{5mm} \forall \vec{t} \in \mathbf{Z}_p^n.
\end{eqnarray*}
\qed

\begin{corollary}
\label{col-encoding}
For any $U_e \in {\cal M}_g(S)$, we have $U_e \in N({\cal P}_n)$.
\end{corollary}
\startproof
Since $\{ \mathsf{X}^n(\vec{s}) \}$ and 
$\{ \mathsf{Z}^n(\vec{t}) \}$ are the generator set
of ${\cal P}_n$ for $p \ge 3$, and
$\{ \mathsf{X}^n(\vec{s}) \}$ and 
$\{ \mathsf{Z}^n(\vec{t}) \}$ and $\mathbf{i} I_{p^n}$
are the generator set of ${\cal P}_n$ for $p=2$, from Lemma \ref{encoded-x-z},
we have $U_e \in N({\cal P}_n)$ \qed

\begin{lemma}
\label{encoding-c-g}
For $U_c \in {\cal M}_{\mathrm{cl}}(S)$, there exists
$U_e \in {\cal M}_g(S)$ such that $U_c = U_e$.
\end{lemma}
\startproof
We will construct $U_e \in {\cal M}_g(S)$
such that $U_e = U_c$.
We set $\ket{\psi}$ by
\begin{eqnarray*}
\ket{\psi} &=& U_c \ket{\vec{0}},
\end{eqnarray*}
and set $\vec{\xi}_{n-k+1},\ldots,\vec{\xi}_n$ and $\theta_z(\cdot)$ by 
\begin{eqnarray}
\widetilde{\mathsf{Z}}^n(\vec{f}_i) &=& \theta_z(\vec{f}_i)
\mathsf{XZ}^n(\vec{\xi}_i) 
\hspace{3mm} \mbox{for } i = 1,\ldots,n-k 
\label{def-z-2} \\ 
\widetilde{\mathsf{Z}}^n(\vec{f}_i) &=& \theta_z(\vec{f}_i)
\mathsf{XZ}^n(\vec{\xi}_i)
= U_c \mathsf{Z}^n(\vec{f}_i) U_c^* \nonumber \\
&& \hspace{10mm} \mbox{for } i=n-k+1,\ldots,n,
\label{def-z}
\end{eqnarray}
for $p \ge 3$, and
\begin{eqnarray}
\label{z-def-2-2}
\widetilde{\mathsf{Z}}^n(\vec{f}_j) &=&
\theta_z(\vec{f}_j) \mu(\vec{\xi}_j)\mathsf{XZ}^n(\vec{\xi}_j)
\hspace{3mm} \mbox{for } j=1,\ldots,n-k, \\
\label{z-def-2}
\widetilde{\mathsf{Z}}^n(\vec{f}_j) &=&
\theta_z(\vec{f}_j) \mu(\vec{\xi}_j) \mathsf{XZ}(\vec{\xi}_j) =
U_c \mathsf{Z}^n(\vec{f}_j) U_c^* \nonumber \\
&& \hspace{10mm} \mbox{for } j=n-k+1,\ldots,n, 
\label{x-def-2}
\end{eqnarray}
for $p=2$,
where $\vec{f}_i$ is a vector such that the $i$-th element is $1$ and
the other elements are $0$.
Note that $\vec{\xi}_{n-k+1},\ldots,\vec{\xi}_n$ are determined 
by $U_c$, while $\vec{\xi}_1,\ldots,\vec{\xi}_{n-k}$
are fixed basis of $C$ 
as we stated in Section \ref{preliminary}.
From  Definition \ref{definition-of-encoding},
we have $\ket{\psi} \in Q(\vec{0})$.
For $i=1,\ldots,n-k$, we set $\theta_z(\vec{f}_i)$ so that
$\widetilde{\mathsf{Z}}^n(\vec{f}_i) \ket{\psi} = \ket{\psi}$.
Specifically, since $Q(\vec{0})$ is an eigenspace that belongs
to an eigenvalue $\lambda_i$ of $\mathsf{XZ}^n(\vec{\xi}_i)$ for
$i= 1,\ldots,n-k$,
we set $\theta_z(\vec{f}_i) = \overline{\lambda}_i$ for
$p \ge 3$ and 
$\theta_z(\vec{f}_i)=\overline{\lambda}_i \overline{\mu(\vec{\xi}_i)}$
for $p=2$.

Set $\vec{\eta}_1,\ldots,\vec{\eta}_n$, and $\theta_x(\cdot)$ by
\begin{eqnarray}
\widetilde{\mathsf{X}}^n(\vec{f}_i) = \theta_x(\vec{f}_i)
\mathsf{XZ}^n(\vec{\eta}_i) 
= U_c \mathsf{X}^n(\vec{f}_i) U_c^*
\hspace{3mm} \mbox{for } i=1,\ldots,n,
\label{def-x}
\end{eqnarray}
for $p \ge 3$, and
\begin{eqnarray}
\widetilde{\mathsf{X}}^n(\vec{f}_j) &=&
\theta_x(\vec{f}_j) \mu(\vec{\eta}_j) \mathsf{XZ}(\vec{\eta}_j) =
U_c \mathsf{X}^n(\vec{f}_j) U_c^*
\hspace{3mm} \mbox{for } j=1,\ldots,n
\end{eqnarray}
for $p=2$.
Then we have
\begin{eqnarray*}
\widetilde{\mathsf{X}}^n(\vec{u}) =
\prod_{i=1}^n \left( \widetilde{\mathsf{X}}^n(\vec{f}_i)\right)^{u_i}
= U_c \mathsf{X}^n(\vec{u}) U_c^* 
\hspace{3mm} \mbox{for } \vec{u} \in \mathbf{Z}_p^n.
\end{eqnarray*}
We also have 
\begin{eqnarray*}
U_c \ket{\vec{u}} &=& U_c \mathsf{X}^n(\vec{u}) \ket{\vec{0}} \\
&=& U_c \mathsf{X}^n(\vec{u}) U_c^* U_c \ket{\vec{0}} \\
&=& \widetilde{\mathsf{X}}^n(\vec{u}) \ket{\psi}.
\end{eqnarray*}

Next, we show 
\begin{eqnarray}
\widetilde{\mathsf{Z}}^n(\vec{f}_i) =
U_c \mathsf{Z}^n(\vec{f}_i) U_c^*
\hspace{3mm} \mbox{for }i=1,\ldots,n-k.
\label{encoded-z-1-n-k}
\end{eqnarray}
Let $\vec{u} = (e_1,\ldots,e_{n-k},x_1,\ldots,x_k) \in \mathbf{Z}_p^n$,
$\vec{e} = (e_1,\ldots,e_{n-k})$,
and $\ket{\varphi(\vec{u})} = U_c \ket{\vec{u}}$.
Then, since $\ket{\varphi(\vec{u})} \in Q(\vec{e})$
and $\mathsf{XZ}^n(\vec{\xi}_i) \ket{\varphi(\vec{u})} =
\lambda_i \omega^{e_i} \ket{\varphi(\vec{u})}$,
we have 
\begin{eqnarray*}
\widetilde{\mathsf{Z}}^n(\vec{f}_i) \ket{\varphi(\vec{u})} &=&
\theta_z(\vec{f}_i) \mathsf{XZ}^n(\vec{\xi}_i) \ket{\varphi(\vec{u})} \\
&=& \omega^{e_i} \ket{\varphi(\vec{u})} \\
&=& \omega^{e_i} U_c \mathsf{X}^n(\vec{u}) U_c^* \ket{\psi} \\
&=& \omega^{e_i} U_c \mathsf{X}^n(\vec{u}) U_c^*
U_c \mathsf{Z}^n(\vec{f}_i) \ket{\vec{0}} \\
&=& \omega^{e_i} U_c \mathsf{X}^n(\vec{u}) U_c^*
U_c \mathsf{Z}^n(\vec{f}_i) U_c^* \ket{\psi} \\
&=& \omega^{e_i} \omega^{-e_i} U_c \mathsf{Z}^n(\vec{f}_i) U_c^*
U_c \mathsf{X}^n(\vec{u}) U_c^* \ket{\psi} \\
&=& U_c \mathsf{Z}^n(\vec{f}_i) U_c^* \ket{\varphi(\vec{u})},
\end{eqnarray*}
for $i= 1,\ldots,n-k$.
Since $\{ \ket{\varphi(\vec{u})} \mid \vec{u} \in \mathbf{Z}_p^n\}$
form an orthonormal basis of ${\cal H}^{\otimes n}$,
Eq.~(\ref{encoded-z-1-n-k}) is satisfied.

From Eqs.~(\ref{relation-xz}), (\ref{def-z-2}),
(\ref{def-z}), (\ref{def-x}) and (\ref{encoded-z-1-n-k}),
we have
\begin{eqnarray*}
\mathsf{XZ}^n(\vec{\xi}_i) \mathsf{XZ}^n(\vec{\xi}_j) 
&=& \mathsf{XZ}^n(\vec{\xi}_j) \mathsf{XZ}^n(\vec{\xi}_i) \\
\mathsf{XZ}^n(\vec{\eta}_i) \mathsf{XZ}^n(\vec{\eta}_j)
&=& \mathsf{XZ}^n(\vec{\eta}_j) \mathsf{XZ}^n(\vec{\eta}_i) \\
\mathsf{XZ}^n(\vec{\xi}_i) \mathsf{XZ}^n(\vec{\eta}_i) 
&=& \omega \mathsf{XZ}^n(\vec{\eta}_i) \mathsf{XZ}^n(\vec{\xi}_i) \\
\mathsf{XZ}^n(\vec{\xi}_i) \mathsf{XZ}^n(\vec{\eta}_j) 
&=& \mathsf{XZ}^n(\vec{\eta}_j) \mathsf{XZ}^n(\vec{\xi}_i),
\end{eqnarray*}
which mean that 
$\vec{\xi}_1,\ldots,\vec{\xi}_n$ and
$\vec{\eta}_1,\ldots,\vec{\eta}_n$ satisfy
Eq.~(\ref{hyperboric}).
It is easy to check that 
$\ket{\psi}$ is an eigenvector of
$\mathsf{XZ}^n(\vec{\xi}_1),\ldots,\mathsf{XZ}^n(\vec{\xi}_n)$,
thus we can write $\ket{\psi} = \ket{\psi(\vec{0})} \in Q_{\min}(\vec{0})$
for some $Q_{\min}(\vec{0})$.

Consequently, we can construct an encoding operator
$U_e \in {\cal M}_g(S)$ such that $U_e = U_c$. \qed

From Corollary \ref{col-encoding} and Lemma \ref{encoding-c-g}, 
we have the following theorem.

\begin{theorem}
\label{theorem-enc}
For a given stabilizer $S$, ${\cal M}_g(S) = {\cal M}_{\mathrm{cl}}(S)$.
\end{theorem}

\subsection{Classification of encoding operators}
\label{classification}

In this section, we show the correspondence between
Bell states and Bell states encoded by encoding operators
(Lemma \ref{lemma-encoded-Bell}, Corollary \ref{corollary-encoded-Bell},
and Corollary \ref{corollary-inverse}).
Then we show the output state  of our EDPs is
always a probabilistic mixture of Bell states if
the input state of protocols is a probabilistic mixture of
Bell states (Theorem \ref{theorem-mixture}). 
Then,
we classify encoding operators into equivalence classes such
that EDPs constructed from encoding operators in the same equivalence class
have the same performance when the input of EDPs are
the probabilistic mixture of Bell states
(Definition \ref{definition_similar}, and Theorems \ref{performance-equal},
\ref{performance-equal-2}).

\begin{lemma}
\label{lemma-encoded-Bell}
The Bell state $\ket{\beta^k(\vec{0})}$ with 
ancilla qubits $\ket{\vec{e}}_A \otimes \ket{\vec{e}}_B$, i.e.,
\begin{eqnarray*}
\ket{\beta^k(\vec{0}),\vec{e}} =
\frac{1}{\sqrt{p^k}} \sum_{\vec{v} \in \mathbf{Z}_p^k}
\ket{\vec{e}}_A \otimes \ket{\vec{v}}_A 
\otimes \ket{\vec{e}}_B \otimes \ket{\vec{v}}_B,
\end{eqnarray*} 
is mapped by 
$\overline{U_e} \otimes U_e$ to
\begin{eqnarray}
\ket{\phi(\vec{e})} = \frac{1}{\sqrt{p^k}}
\sum_{\vec{u} \in \vec{e} \times \mathbf{Z}_p^k}
\overline{\widetilde{\mathsf{X}}}^n(\vec{u}) \overline{\ket{\psi(\vec{0})}}
\otimes \widetilde{\mathsf{X}}(\vec{u}) \ket{\psi(\vec{0})},
\end{eqnarray}
where $\overline{\widetilde{\mathsf{X}}}^n(\vec{u})$ is
the complex conjugated matrix of $\widetilde{\mathsf{X}}(\vec{u})$,
and $\vec{e} \times \mathbf{Z}_p^k$ is the subset
$\{ (e_1,\ldots,e_{n-k},x_1,\ldots,x_k) \mid x_i \in \mathbf{Z}_p \}$
of $\mathbf{Z}_p^n$.
\end{lemma}

\startproof
It is obvious from Eq.~(\ref{def-encoding-operator}) in 
the definition of the encoding operator 
$U_e$. \qed

\begin{corollary}
\label{corollary-encoded-Bell}
A Bell state
\begin{eqnarray}
\label{errored-Bell}
I_{p^k} \otimes \mathsf{X}^k(\vec{\ell})
\mathsf{Z}^k(\vec{m}) \ket{\beta^k(\vec{0})}
\end{eqnarray}
with ancilla qubits $\ket{\vec{e}}_A \otimes \ket{\vec{e}}_B$, i.e.,
\begin{eqnarray}
\label{errored-Bell-ancilla}
\ket{\beta^k(\vec{\ell},\vec{m}),\vec{e}} =
\frac{1}{\sqrt{p^k}} \sum_{\vec{v} \in \mathbf{Z}_p^k}
\ket{\vec{e}}_A \otimes \ket{\vec{v}}_A 
\otimes \ket{\vec{e}}_B \otimes 
\mathsf{X}^k(\vec{\ell})\mathsf{Z}^k(\vec{m})\ket{\vec{v}}_B,
\end{eqnarray} 
is mapped by $\overline{U_e} \otimes U_e$ to
\begin{eqnarray}
\label{encoded-Bell}
I_{p^n} \otimes \mathsf{XZ}^n(\vec{\ell}G+\vec{m}H) \ket{\phi(\vec{e})},
\end{eqnarray}
multiplied by a scalar of unit absolute value,
where the matrices $G$ and $H$ are
\begin{eqnarray*}
G = \left(\begin{array}{c}
\vec{\eta}_{n-k+1} \\ \vdots \\ \vec{\eta}_n
\end{array}\right),  \hspace{5mm}
H = \left(\begin{array}{c}
\vec{\xi}_{n-k+1} \\ \vdots \\ \vec{\xi}_n
\end{array}\right).
\end{eqnarray*}
\end{corollary}

\startproof
Let 
\begin{eqnarray}
\label{ell-prime}
\vec{\ell}^\prime &=& (0,\ldots,0,\ell_1,\ldots,\ell_k) \in \mathbf{Z}_p^n \\
\label{m-prime}
\vec{m}^\prime &=& (0,\ldots,0,m_1,\ldots,m_k) \in \mathbf{Z}_p^n.
\end{eqnarray}
From Eqs.~(\ref{eq-uxu}) and (\ref{eq-uzu}),
\begin{eqnarray*}
U_e \mathsf{X}^n(\vec{\ell}^\prime)\mathsf{Z}^n(\vec{m}^\prime) U_e^*
= \widetilde{\mathsf{X}}^n(\vec{\ell}^\prime)
\widetilde{\mathsf{Z}}^n(\vec{m}^\prime).
\end{eqnarray*}
Thus, a state in Eq.~(\ref{errored-Bell-ancilla}) is mapped by
$\overline{U_e} \otimes U_e$ to
\begin{eqnarray*}
&&(\overline{U_e} \otimes U_e) \ket{\beta^k(\vec{\ell},\vec{m}),\vec{e}} \\
&=& (\overline{U_e} \otimes U_e)
(I_{p^n} \otimes 
\mathsf{X}^n(\vec{\ell}^\prime)\mathsf{Z}^n(\vec{m}^\prime))
\ket{\beta^k(\vec{0}),\vec{e}} \\
&=& (\overline{U_e} \otimes U_e) (I_{p^n} \otimes 
\mathsf{X}^n(\vec{\ell}^\prime)\mathsf{Z}^n(\vec{m}^\prime))
(\overline{U_e}^* \otimes U_e^*)(\overline{U_e} \otimes U_e)
\ket{\beta^k(\vec{0}),\vec{e}} \\
&=& I_{p^n} \otimes \widetilde{\mathsf{X}}^n(\vec{\ell}^\prime)
\widetilde{\mathsf{Z}}^n(\vec{m}^\prime) \ket{\phi(\vec{e})} \\
&\stackrel{{\tiny (a)}}{\simeq}& I_{p^n} \otimes \mathsf{XZ}^n(\vec{\ell}G)
\mathsf{XZ}^n(\vec{m}H) \ket{\phi(\vec{e})} \\
&\simeq& I_{p^n} \otimes \mathsf{XZ}^n(\vec{\ell}G+\vec{m}H) 
\ket{\phi(\vec{e})},
\end{eqnarray*}
where $\simeq$ denotes that one vector is equal to another vector
multiplied by a scalar of unit absolute value.
Note that (a) follows from Eqs. (\ref{relation-xz}),
(\ref{encoded-x}), and (\ref{encoded-z}).
\qed

\begin{corollary}
\label{corollary-inverse}
The state 
\begin{eqnarray*}
I_{p^n} \otimes \mathsf{XZ}^n(\vec{\ell}G+\vec{m}H) \ket{\phi(\vec{e})}
\end{eqnarray*}
is mapped by $\overline{U_e}^* \otimes U_e^*$ to
\begin{eqnarray*}
\ket{\beta^k(\vec{\ell},\vec{m}),\vec{e}} =
\frac{1}{\sqrt{p^k}} \sum_{\vec{v} \in \mathbf{Z}_p^k}
\ket{\vec{e}}_A \otimes \ket{\vec{v}}_A 
\otimes \ket{\vec{e}}_B \otimes 
\mathsf{X}^k(\vec{\ell})\mathsf{Z}^k(\vec{m})\ket{\vec{v}}_B
\end{eqnarray*}
multiplied by a scalar of unit absolute value,
i.e., $\ket{\beta^k(\vec{w})}$ with ancilla qudits 
$\ket{\vec{e}}_A \otimes \ket{\vec{e}}_B$,
where $\vec{w} = (\ell_1,\ldots,\ell_k|m_1,\ldots,m_k)$. \qed
\end{corollary}

\begin{definition}
For a vector $\vec{s} = (s_1,\ldots,s_{n-k})$,
we define the set $D(\vec{s})$ by
\begin{eqnarray*}
D(\vec{s}) = \{
\vec{t} \in \mathbf{Z}_p^{2n} \mid 
\langle \vec{\xi}_i, \vec{t} \rangle = s_i \}.
\end{eqnarray*}
\end{definition}

\begin{lemma}
\label{after-error-correction}
When we apply Steps \ref{step1}--\ref{step4} of our distillation
protocol to the state $\ket{\beta^n(\vec{t})}$ and
Alice and Bob  do not abort the protocol in Step \ref{step-abort},
the resulting quantum state is 
\begin{eqnarray*}
I_{p^k} \otimes \mathsf{XZ}^n(f(\vec{t})) \ket{\phi(\vec{a})},
\end{eqnarray*}
where $f(\cdot)$ is the mapping from $\mathbf{Z}_p^{2n}$ to
$C^\bot$ and depends on the error correction process in 
Step \ref{step4}.
Specifically, $f(\cdot)$ is defined as follows.
Let $\vec{t}^\prime$ be the most likely error in $D(\vec{b}-\vec{a})$.
The mapping $f(\cdot)$ is defined as
\begin{eqnarray}
\label{def-f}
f:~D(\vec{b}-\vec{a}) \ni \vec{x} 
\mapsto \vec{x} - \vec{t}^\prime \in C^\bot
\end{eqnarray}
for each $D(\vec{b}-\vec{a})$. Note that
$\cup_{\vec{s} \in \mathbf{Z}_p^{n-k}}D(\vec{s})= \mathbf{Z}_p^{2n}$.
\end{lemma}

\startproof
After Steps \ref{step1} and \ref{step2},
the state becomes 
\begin{eqnarray*}
\mathsf{P}^\star(\vec{a}) \otimes \mathsf{P}(\vec{b})
\ket{\beta^n(\vec{t})} = 
I_{p^k} \otimes \mathsf{XZ}^n(\vec{t}) \ket{\phi(\vec{a})}
\in Q^\star(\vec{a}) \otimes Q(\vec{b}),
\end{eqnarray*}
where $\mathsf{P}^\star(\vec{a})$ and $\mathsf{P}(\vec{b})$
represent the projection on to $Q^\star(\vec{a})$ and
$Q(\vec{b})$ respectively.
In Step \ref{step4}, Bob decides the most likely error 
$\vec{t}^\prime \in D(\vec{b}-\vec{a})$ and applies 
$M = \mathsf{XZ}^n(-\vec{t}^\prime)$.
Then the state becomes 
\begin{eqnarray*}
I_{p^k} \otimes \mathsf{XZ}^n(\vec{t}-\vec{t}^\prime) \ket{\phi(\vec{a})}
= I_{p^k} \otimes \mathsf{XZ}^n(f(\vec{t})) \ket{\phi(\vec{a})}
\in Q^\star(\vec{a}) \otimes Q(\vec{a}).
\end{eqnarray*}
The condition  $M Q(\vec{b}) = Q(\vec{a})$ implies
$\vec{t}-\vec{t}^\prime \in C^\bot$.
\qed

\begin{remark}
\label{remark-f}
The mapping $f(\cdot)$ does not depends on the choice of
a basis $\{ \vec{\xi}_1,\ldots,\vec{\xi}_{n-k}\}$ of
$C$ or the joint eigenspace $Q(\vec{0})$.
Since there exists one to one correspondence between
$D(\vec{s})$ and a coset of $\mathbf{Z}_p^{2n}/C^\bot$,
the mapping $f(\cdot)$ is defined only by a representative $\vec{t}^\prime$
of each coset of $\mathbf{Z}_p^{2n}/C^\bot$ in Eq.~(\ref{def-f}).
\end{remark}

\begin{lemma}
When we apply Step \ref{step1}--\ref{step6} of our distillation
protocol to the state $\ket{\beta^n(\vec{t})}$ and
Alice and Bob do not abort the protocol in Step \ref{step-abort},
the resulting quantum state is 
\begin{eqnarray*}
\ket{\beta^k(\vec{w})} = \ket{\beta^k(g \circ f (\vec{t}))},
\end{eqnarray*}
where the mapping $g$ is the mapping from $C^\bot$ to $\mathbf{Z}_p^{2k}$,
more precisely  
\begin{eqnarray}
g:~C^\bot \ni \vec{\ell}G+\vec{m}H+\vec{v} \mapsto \vec{w} = 
(\ell_1,\ldots,\ell_k|m_1,\ldots,m_k) \in \mathbf{Z}_p^{2k}
\hspace{5mm} \forall~\vec{v} \in C.
\label{map-g}
\end{eqnarray}
\end{lemma}
\startproof
From Lemma \ref{after-error-correction},
after Steps \ref{step1}--\ref{step4}
the resulting quantum state is 
\begin{eqnarray*}
I_{p^k} \otimes \mathsf{XZ}^n(f(\vec{t})) \ket{\phi(\vec{a})},
\end{eqnarray*}
with $f(\vec{t}) \in C^\bot$.
Since $\vec{\xi}_1,\ldots,\vec{\xi}_n,
\vec{\eta}_{n-k+1},\ldots,\vec{\eta}_n$ form
a basis of $C^\bot$
and $\vec{\xi}_1,\ldots,\vec{\xi}_{n-k}$ form a basis of $C$,
$f(\vec{t})$ can be written as a linear
combination
\begin{eqnarray}
\label{decomposition-c-perp}
f(\vec{t}) = \sum_{i=1}^k \ell_i \vec{\eta}_{n-k+i} 
+ m_i \vec{\xi}_{n-k+i} + \vec{v},
\end{eqnarray} 
where $\vec{v} \in C$.
Since $\ket{\phi(\vec{a})}$ is a joint engeinvector of $S$, 
\begin{eqnarray*}
I_{p^k} \otimes \mathsf{XZ}^n(f(\vec{t})) \ket{\phi(\vec{a})}
&=& I_{p^n} \otimes \mathsf{XZ}^n(\vec{\ell}G+\vec{m}H + \vec{v}) 
\ket{\phi(\vec{a})} \\
&\simeq&  I_{p^n} \otimes \mathsf{XZ}^n(\vec{\ell}G+\vec{m}H) 
\ket{\phi(\vec{a})}
\end{eqnarray*}
By Corollary \ref{corollary-inverse},
after Step \ref{step5} and \ref{step6}
the quantum state becomes 
$\ket{\beta^k(\vec{w})} = \ket{\beta^k(g \circ f(\vec{t}))}$,
where $\vec{w} = (\ell_1,\ldots,\ell_k|m_1,\ldots,m_k)$.
\qed
\begin{theorem}
\label{theorem-mixture}
When the input to our distillation protocol is a
probabilistic mixture of Bell states 
$\ket{\beta^n(\vec{t})}$ for $\vec{t} \in \mathbf{Z}_p^{2n}$, i.e.,
\begin{eqnarray}
\rho_{in} = \sum_{\vec{t} \in \mathbf{Z}_p^{2n}} P_{in}(\vec{t}) 
\ket{\beta^n(\vec{t})}\bra{\beta^n(\vec{t})}
\label{pro-mixture}
\end{eqnarray}
and the difference of Alice and Bobs' measurement result
is $\vec{b}-\vec{a} \in T$,
then the output from our distillation protocol is also
probabilistic mixture of Bell states
$\ket{\beta^k(\vec{w})}$ for $\vec{w} \in \mathbf{Z}_p^{2k}$, i.e.,
\begin{eqnarray*}
\rho_{out} =
\sum_{\vec{w} \in \mathbf{Z}_p^{2k}} P_{out}(\vec{w}) 
\ket{\beta^k(\vec{w})}\bra{\beta^k(\vec{w})},
\end{eqnarray*}
where $P_{out}(\vec{w})$ is given by
\begin{eqnarray}
P_{out}(\vec{w}) = 
\sum_{\vec{t} \in D(\vec{b}-\vec{a}):
g \circ f(\vec{t}) = \vec{w}}
P_{in}^\prime(\vec{t}),
\label{p-out}
\end{eqnarray}
and $P_{in}^\prime(\vec{t})$ is normalized as
\begin{eqnarray*}
P_{in}^\prime(\vec{t}) = \frac{P_{in}(\vec{t})}{
\sum_{\vec{t} \in D(\vec{b}-\vec{a})} P_{in}(\vec{t})}.
\end{eqnarray*}
\end{theorem}
\startproof
After Steps 1--4 of our distillation protocol,
from the linearity of the measurement and the error 
correction, the input state $\rho_{in}$ becomes 
\begin{eqnarray*}
\rho^\prime =
\sum_{\vec{t} \in D(\vec{b}-\vec{a})}
P_{in}^\prime(\vec{t}) \left(
I_{p^k} \otimes \mathsf{XZ}^n(f(\vec{t})) \ket{\phi(\vec{a})}
\bra{\phi(\vec{e})} I_{p^k} \otimes \mathsf{XZ}^n(f(\vec{t}))^* \right).
\end{eqnarray*}
After applying the inverse of the encoding operator,
the state $\rho^\prime$ becomes 
\begin{eqnarray}
\label{eq-coefficient-1}
\rho_{out} &=&
\sum_{\vec{t} \in D(\vec{b}-\vec{a})}
P_{in}^\prime(\vec{t}) \ket{\beta^k(g\circ f(\vec{t}))}
\bra{\beta^k(g \circ f(\vec{t}))} \\
\label{eq-coefficient-2}
&=& \sum_{\vec{w} \in \mathbf{Z}_p^{2k}} P_{out}(\vec{w})
\ket{\beta^k(\vec{w})}\bra{\beta^k(\vec{w})},
\end{eqnarray} 
where $P_{out}(\vec{w})$ is given by
\begin{eqnarray*}
P_{out}(\vec{w}) = 
\sum_{\vec{t} \in D(\vec{b}-\vec{a}):
g \circ f(\vec{t}) = \vec{w}}
P_{in}^\prime(\vec{t}).
\end{eqnarray*}
\qed

When the input of EDPs are the probabilistic mixture of Bell states,
the performance of the distillation protocol only depends on the
coefficients $P_{out}(\vec{w})$ of the output of the protocol.
Hereafter, we fix the stabilizer $S$ and the error correction
process $f(\cdot)$.

\begin{definition}
\label{definition_similar}
For two stabilizer based EDPs constructed from encoding operators
$U_e$ and $V_e$ respectively,
let the mapping $g_U$ be determined by $U_e$ in Eq.~(\ref{map-g})
and $g_V$ be determined by $V_e$ in Eq.~(\ref{map-g}).
If $g_U(\cdot) = g_V(\cdot)$, then we define
two encoding operators $U_e$ and $V_e$ are similar
and denote it by $U_e \sim V_e$.
\end{definition} 

\begin{theorem}
\label{performance-equal}
For two stabilizer based EDPs constructed from encoding operators
$U_e$ and $V_e$ respectively, let 
\begin{eqnarray*}
\rho_{out,U_e} = \sum_{\vec{w} \in \mathbf{Z}_p^{2k}} P_{out,U_e}(\vec{w})
\ket{\beta^k(\vec{w})}\bra{\beta^k(\vec{w})}
\end{eqnarray*}
and 
\begin{eqnarray*}
\rho_{out,V_e} = \sum_{\vec{w} \in \mathbf{Z}_p^{2k}} P_{out,V_e}(\vec{w})
\ket{\beta^k(\vec{w})}\bra{\beta^k(\vec{w})}
\end{eqnarray*}
be output states of each protocols when inputs of each protocol
are Eq.~(\ref{pro-mixture}).
If $U_e \sim V_e$, then we have
\begin{eqnarray}
\label{p-out-equal}
P_{out,U_e}(\vec{w}) = P_{out,V_e}(\vec{w}) \hspace{5mm}
\forall~\vec{w} \in \mathbf{Z}_p^{2k},
\end{eqnarray}
i.e., performances of two protocols are the same.
\end{theorem} 
\startproof
From Eq.~(\ref{p-out}) and the fact that $g_U(\cdot) = g_V(\cdot)$,
for any $\vec{w} \in \mathbf{Z}_p^{2k}$
\begin{eqnarray*}
P_{out,U_e}(\vec{w}) &=& \sum_{\vec{t} \in D(\vec{b}-\vec{a}):
g_U \circ f(\vec{t}) = \vec{w}} P_{in}^\prime(\vec{t}) \\
&=& \sum_{\vec{t} \in D(\vec{b}-\vec{a}):
g_V \circ f(\vec{t}) = \vec{w}} P_{in}^\prime(\vec{t}) = P_{out,V_e}(\vec{w}).
\end{eqnarray*}
\qed

\begin{theorem}
\label{performance-equal-2}
If Eq.~(\ref{p-out-equal}) holds for any input state of the form
in Eq.~(\ref{pro-mixture}), then
$U_e \sim V_e$.
\end{theorem}
\startproof
We prove the contraposition of this statement, i.e.,
if $U_e \not\sim V_e$, then Eq.~(\ref{p-out-equal}) does
not hold for some input states.
Since $g_U(\cdot) \neq g_V(\cdot)$,
there exists $\vec{u} \in C^\bot$ such that $g_U(\vec{u}) \neq g_V(\vec{u})$.
Consider the following  input state.
Let 
\begin{eqnarray*}
P_{in}(\vec{t}) = \left\{ \begin{array}{ll}
\frac{1}{|\{ \vec{s} \in \mathbf{Z}_p^{2n}  \mid f(\vec{s})=\vec{u} \}|}
 & \mbox{if } f(\vec{t}) = \vec{u} \\
0 & \mbox{if } f(\vec{t}) \neq \vec{u}
\end{array} \right. .
\end{eqnarray*}
Then we have
\begin{eqnarray*}
P_{out, U_e}(\vec{w}) &=& \left\{ \begin{array}{ll}
1 & \mbox{if } \vec{w} = g_U(\vec{u}) \\
0 & \mbox{if } \vec{w} \neq g_U(\vec{u})
\end{array} \right. , \\ 
P_{out, V_e}(\vec{w}) &=& \left\{ \begin{array}{ll}
1 & \mbox{if } \vec{w} = g_V(\vec{u}) \\
0 & \mbox{if } \vec{w} \neq g_V(\vec{u})
\end{array} \right. ,
\end{eqnarray*}
and Eq.~(\ref{p-out-equal}) does not holds. \qed

\subsection{Enumeration of equivalence classes of encoding operators}

Classify ${\cal M}_g(S)$ into equivalence classes by $\sim$,
and denote the representative set of the equivalence classes by 
$\widehat{{\cal M}}_g(S)$. 
In this section, we show how to enumerate all elements of
$\widehat{{\cal M}}_g(S)$ in Theorem \ref{theorem-1}.

\begin{lemma} 
\label{lemma-classify-2}
Let two encoding operators $U_e$ and $V_e$ be
constructed from $\{ \vec{\xi}_{n-k+1},\ldots,\vec{\xi}_n$,
$\vec{\eta}_{n-k+1},\ldots,\vec{\eta}_n \}$
and $\{ \vec{\xi}_{n-k+1}^\prime,\ldots,\vec{\xi}_n^\prime$,
$\vec{\eta}_{n-k+1}^\prime,\ldots,\vec{\eta}_n^\prime \}$
respectively and the other parameters 
(a)  $\theta_x(\cdot)$,
(b)  $\vec{\eta}_1,\ldots,\vec{\eta}_{n-k}$,
and (c)  $Q_{\min}(\vec{0})$ be the same.
Further assume that $\vec{\xi}_i \equiv \vec{\xi}_i^\prime \pmod{C}$ for all
$n-k+1 \le i \le n$ and
$\vec{\eta}_i \equiv \vec{\eta}_i^\prime \pmod{C}$ for all
$n-k+1 \le i \le n$.
Then $U_e \sim V_e$.
\end{lemma}
\startproof
Let $g_U$, $G_U$, and $H_U$ be determined by $U_e$ in Eq.~(\ref{map-g}),
and $g_V$, $G_V$, and $H_V$ be determined by $V_e$ in Eq.~(\ref{map-g}).
For any vector $\vec{u} \in C^\bot$,
we have
\begin{eqnarray*}
\vec{u} = \vec{\ell}G_U + \vec{m} H_U + \vec{v} =
\vec{\ell} G_V + \vec{m} H_V + \vec{v}^\prime \hspace{5mm} 
\exists~\vec{v},~\vec{v}^\prime \in C.
\end{eqnarray*}
Thus, we have 
\begin{eqnarray*}
g_U(\vec{u}) = g_V(\vec{u}) \hspace{5mm} \forall~\vec{u} \in C^\bot.
\end{eqnarray*}
and $U_e \sim V_e$. \qed

\begin{lemma}
\label{lemma-classify-3}
Let two encoding operators $U_e$ and $V_e$ be
constructed from $\{ \vec{\xi}_{n-k+1},\ldots,\vec{\xi}_n$,
$\vec{\eta}_{n-k+1},\ldots,\vec{\eta}_n \}$
and $\{ \vec{\xi}_{n-k+1}^\prime,\ldots,\vec{\xi}_n^\prime$,
$\vec{\eta}_{n-k+1}^\prime,\ldots,\vec{\eta}_n^\prime \}$
respectively and the other parameters 
(a). $\theta_x(\cdot)$,
(b). $\vec{\eta}_1,\ldots,\vec{\eta}_{n-k}$,
and (c). $Q_{\min}(\vec{0})$ are the same.
If $g_U(\cdot) = g_V(\cdot)$, i.e., $U_e \sim V_e$,
then $\vec{\xi}_i \equiv \vec{\xi}_i^\prime \pmod{C}$ for all
$n-k+1 \le i \le n$ and
$\vec{\eta}_i \equiv \vec{\eta}_i^\prime \pmod{C}$ for all
$n-k+1 \le i \le n$.
\end{lemma}
\startproof
For $\vec{u} \in C^\bot$ such that 
\begin{eqnarray*}
g_U(\vec{u}) = g_V(\vec{u}) = (\vec{f}_i|\vec{0}) \in \mathbf{Z}_p^{2k},
\end{eqnarray*}
from Eq.~(\ref{map-g}), we have
\begin{eqnarray*}
\vec{u} = \vec{\xi}_i + \vec{v} = \vec{\xi}_i^\prime + \vec{v}^\prime
\hspace{5mm} \exists~\vec{v},\vec{v}^\prime \in C.
\end{eqnarray*}
Thus, we have
\begin{eqnarray*}
\vec{\xi}_i - \vec{\xi}_i^\prime = \vec{v} - \vec{v}^\prime \in C,
\end{eqnarray*}
which means $\vec{\xi}_i \equiv \vec{\xi}_i^\prime \pmod{C}$ for 
$n-k+1 \le i \le n$.
Similarly, for $\vec{u} \in C^\bot$ such that
\begin{eqnarray*}
g_U(\vec{u}) = g_V(\vec{u}) = (\vec{0}|\vec{f}_i) \in \mathbf{Z}_p^{2k},
\end{eqnarray*}
from Eq.~(\ref{map-g}), we have 
\begin{eqnarray*}
\vec{u} = \vec{\eta}_i + \vec{v} = \vec{\eta}_i^\prime + \vec{v}^\prime
\hspace{5mm} \exists~\vec{v},\vec{v}^\prime \in C.
\end{eqnarray*}
Thus, we have
\begin{eqnarray*}
\vec{\eta}_i - \vec{\eta}_i^\prime = \vec{v} - \vec{v}^\prime \in C,
\end{eqnarray*}
which means $\vec{\eta}_i \equiv \vec{\eta}_i^\prime \pmod{C}$ for 
$n-k+1 \le i \le n$. \qed

\begin{definition}
Let $\vec{x}+C$ and $\vec{y}+C$ be elements of
the coset $C^\bot /C$. Define
a symplectic inner product of $\vec{x}+C$ and $\vec{y}+C$ as
\begin{eqnarray}
\label{coset-inner-product}
\langle \vec{x}+C, \vec{y}+C \rangle = 
\langle \vec{x}, \vec{y} \rangle.
\end{eqnarray}
Note that this inner product does not depend on
choices of a representatives $\vec{x}$ of $\vec{x}+C$ or 
$\vec{y}$ of $\vec{y}+C$.
\end{definition}

\begin{lemma}
The linear space  $C^\bot /C$ is a $2k$--dimensional symplectic
space with respect to the symplectic 
inner product in Eq.~(\ref{coset-inner-product}),
and
$\{ \vec{\xi}_{n-k+1}+C,\ldots,\vec{\xi}_n+C$,
$\vec{\eta}_{n-k+1}+C,\ldots,\vec{\eta}_n+C \}$ form
a hyperbolic basis of $C^\bot / C$.
\end{lemma}
\startproof
It is easy to check that $\{ \vec{\xi}_{n-k+1}+C,\ldots,\vec{\xi}_n+C$,
$\vec{\eta}_{n-k+1}+C,\ldots,\vec{\eta}_n+C \}$ form a basis of $C^\bot / C$.
From Eqs.~(\ref{hyperboric}), we have
\begin{eqnarray*}
\langle \vec{\xi}_i + C , \vec{\eta}_j + C \rangle &=& \delta_{ij}, \\
\langle \vec{\xi}_i + C, \vec{\xi}_j + C \rangle &=& 0 \\
\langle \vec{\eta}_i +C, \vec{\eta}_j + C \rangle &=& 0 
\end{eqnarray*}
for $i,j \in \{n-k+1,\ldots,n\}$. \qed

As a consequence of Lemmas
\ref{lemma-classify-2}, and \ref{lemma-classify-3},
we have the following theorem.

\begin{theorem}
\label{theorem-1}
There is one-to-one correspondence between
Elements of $\widehat{{\cal M}}_g(S)$  and
choices of hyperbolic bases of
$C^\bot / C$ with respect to the inner product in 
Eq.~(\ref{coset-inner-product}).
Specifically, if two encoding operators $U_e$ and $V_e$ are
different only by (a)  $\theta_x(\cdot)$, 
(b) $\vec{\eta}_1,\ldots,\vec{\eta}_{n-k}$, or
(c) $Q_{\min}(\vec{0})$, then $U_e \sim V_e$.
Two encoding operators $U_e$ and $V_e$ are
$U_e \sim V_e$ if and only if
$\vec{\xi}_i \equiv \vec{\xi}_i^\prime \pmod{C}$ 
for all $n-k+1 \le i \le n$ and
$\vec{\eta}_i \equiv \vec{\eta}_i^\prime \pmod{C}$
for all $n-k+1 \le i \le n$, i.e.,
two hyperbolic bases 
$\{ \vec{\xi}_{n-k+1}+C,\ldots,\vec{\xi}_n+C$
$\vec{\eta}_{n-k+1}+C,\ldots,\vec{\eta}_n+C\}$ and
$\{ \vec{\xi}_{n-k+1}^\prime+C,\ldots,\vec{\xi}_n^\prime+C$ 
$\vec{\eta}_{n-k+1}^\prime+C,\ldots,\vec{\eta}_n^\prime+C \}$
of $C^\bot /C$
are equal (Lemmas \ref{lemma-classify-2} and \ref{lemma-classify-3}). 
\end{theorem}

\begin{remark}
When the input of the
protocol is a probabilistic mixture of Bell states,
we can find the best stabilizer based EDP as follows.
For a given parameter $n$ and $k$,
find appropriate values for the following parameters.
\begin{enumerate}
\renewcommand{\theenumi}{\alph{enumi}}
\renewcommand{\labelenumi}{(\theenumi)}
\item a stabilizer $S$: 
a self-orthogonal subspace $C \subset \mathbf{Z}_p^{2n}$.
\item decision rule whether or not to abort the protocol
in Step \ref{step-abort}: a set $T \subset \mathbf{Z}_p^{n-k}$.
\item error correction process: mapping $f(\cdot)$ from 
$\mathbf{Z}_p^{2n}$ to $C^\bot$.
\item an equivalence class of encoding operator:
a hyperbolic basis of $C^\bot/C$.  
\end{enumerate}
\end{remark}

Remark \ref{remark-f} and Theorem \ref{theorem-1}
significantly reduce the number of candidates of good EDPs. 
Indeed, for a given parameter
$n$ and $k$,
we enumerate $n-k$ dimensional self-orthogonal subspaces
$C$ (enumerating stabilizers $S$)
and all hyperbolic bases of $C^\bot /C$ for each $C$
(enumerating the equivalence classes of encoding operators),
instead of all hyperbolic bases of $\mathbf{Z}_p^{2n}$
(enumerating stabilizers $S$ and all encoding operators).
The number of all hyperbolic bases of $\mathbf{Z}_p^{2n}$ is
equal to the cardinality of the set of symplectic mappings on
$\mathbf{Z}_p^{2n}$, i.e.,
$|\mathrm{Sp}_{2n}(\mathbf{Z}_p)| = p^{n^2} \prod_{i=1}^n (p^{2i}-1)$
\cite[Theorem 3.1.2]{symplectic-groups}.
While, the number of $n-k$ dimensional self-orthogonal subspace
of $\mathbf{Z}_p^{2n}$ is 
$\prod_{i=0}^{n-k-1} (p^{2n-i}-p^i)/ (p^{n-k}-p^i)$
(see remark \ref{remark-self-orthogonal}),
and the number of all hyperbolic bases of $C^\bot / C$ is
equal to $|\mathrm{Sp}_{2k}(\mathbf{Z}_p)|=p^{k^2} \prod_{i=1}^k(p^{2i}-1)$.
Thus the number of candidates of EDPs is reduced by
$1/\{p^{n^2-k^2} \prod_{i=1}^{n-k}(p^i-1)\}$.
For example, the number of candidates of EDPs is
reduced by $1/12288$ when $n=4$, $k=2$, and $p=2$.
Note that the number of permutation based EDPs \cite{dehaene03a}
for a given parameter $n$, $k$, and $p$ is 
also same as the number of 
all hyperbolic bases of $\mathbf{Z}_p^{2n}$.

\begin{remark}
\label{remark-self-orthogonal}
The number of $n-k$ dimensional self-orthogonal subspace of 
$\mathbf{Z}_p^{2n}$ is
the number of $n-k$ mutually orthonormal vectors
$\prod_{i=0}^{n-k-1} (p^{2n-i}-p^i)$ divided by
the number of bases of $n-k$ dimensional self-orthogonal subspace
$\prod_{i=0}^{n-k-1} (p^{n-k}-p^i)$.
\end{remark}


\section{EDP with good performance}
\label{good-edp}

\begin{figure}[t]
    \includegraphics[width=\linewidth]{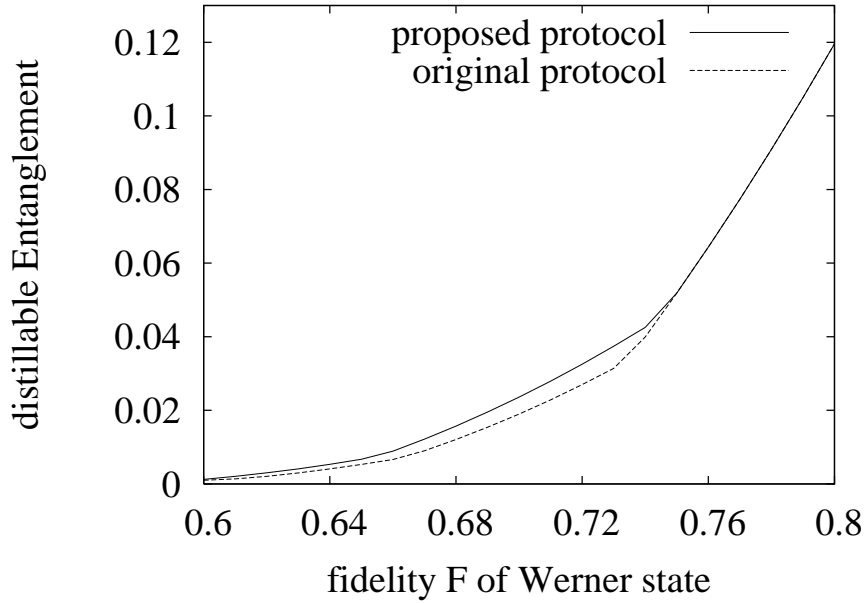}
\vspace{-7mm}
\caption{Comparison of the performance between
the proposed protocol and the protocol 
originally proposed in \cite{matsumoto-epp}.}
\label{graph1}
\end{figure}

\begin{figure}[h]
\vspace{-5mm}
    \includegraphics[width=\linewidth]{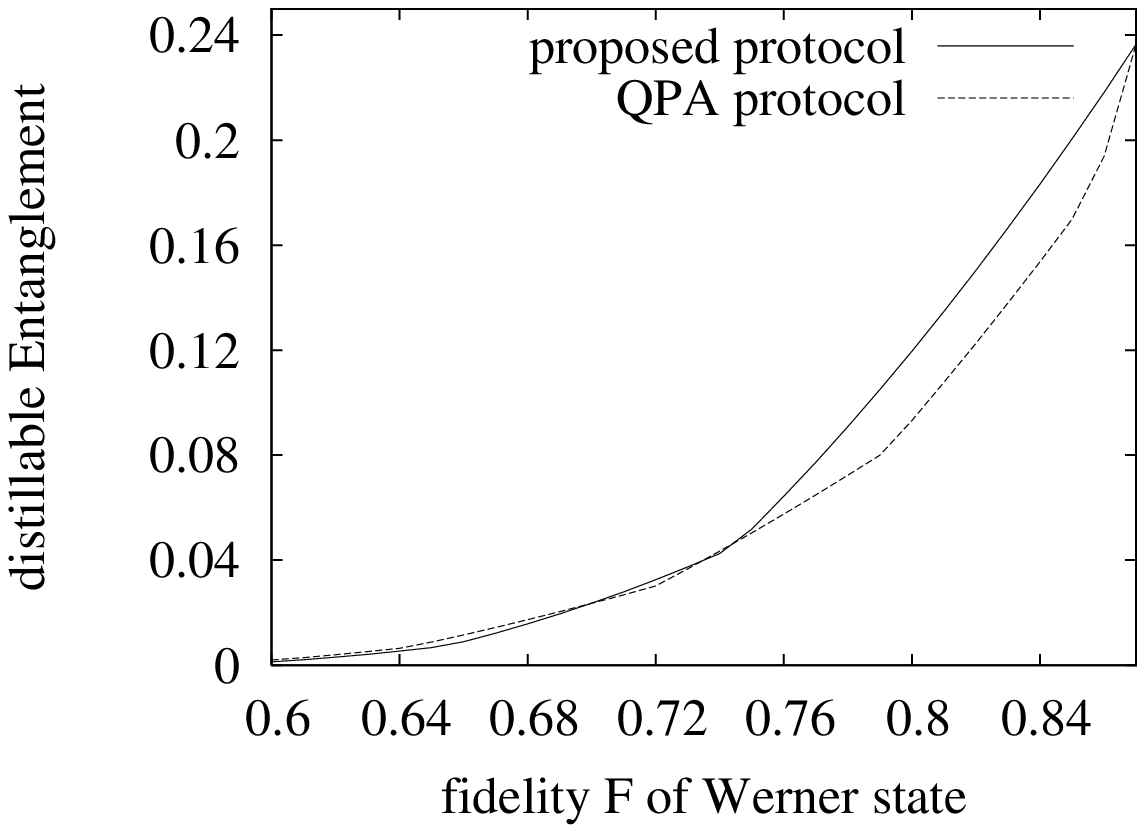}
\vspace{-7mm}
\caption{Comparison of the performance between 
the proposed protocol and the QPA protocol.}
\label{graph2}
\end{figure}

We can improve the performance of the protocol proposed  
in \cite{matsumoto-epp} by choosing an optimal encoding 
operator.
The improved protocol has the 
best performance over the range of fidelity greater than $0.6$
for a parameter
$n=4, k=2$, $p=2$, and $T=\{\vec{0} \}$.
Note that there is no choice of error correction process
when $T = \{ \vec{0} \}$.
We calculated the performance 
by using the protocol appropriate times iteratively
followed by the hashing protocol.
The performance is plotted in
Fig.~\ref{graph1} and is compared to the 
performance of the protocol in \cite{matsumoto-epp}.
The proposed protocol is also compared to the performance
of the QPA protocol in Fig.~\ref{graph2}, and has a better performance than
the QPA protocol over the wide range of fidelity.
We remark that the QPA protocol has the best performance
among EDPs constructed from $[[2,1]]$ stabilizer codes. 

The proposed protocol is constructed from a stabilizer code
with a stabilizer
\begin{eqnarray*}
S = \left\{ X\otimes X \otimes X \otimes X,~
Z \otimes Z \otimes Z \otimes Z \right\}.
\end{eqnarray*}
The encoding operator is constructed as follows.
The vector representation of the stabilizer is
\begin{eqnarray*}
\begin{array}{ll}
\vec{\xi}_1 = (1111|0000), & \vec{\xi}_2 = (0000|1111).
\end{array}
\end{eqnarray*}
Then we choose $\vec{\xi}_3,\vec{\xi}_4$ and
$\vec{\eta}_1,\ldots,\vec{\eta}_4$ to be
\begin{eqnarray*}
\begin{array}{ll}
\vec{\xi}_3 = (1100|0000), & \vec{\xi}_4=(1010|0000), \\
\vec{\eta}_1 = (0000|1110), & \vec{\eta}_2=(1110|0000),\\
\vec{\eta}_3 = (0000|1010), & \vec{\eta}_4=(1010|1100).
\end{array}
\end{eqnarray*}
We choose
\begin{eqnarray*}
\begin{array}{ll}
\widetilde{\mathsf{X}}^4(\vec{f}_1) = Z \otimes Z \otimes Z \otimes I_2 &
\widetilde{\mathsf{X}}^4(\vec{f}_2) = X \otimes X \otimes X \otimes I_2 \\
\widetilde{\mathsf{X}}^4(\vec{f}_3) = Z \otimes I_2 \otimes Z \otimes I_2 &
\widetilde{\mathsf{X}}^4(\vec{f}_4) = \mathbf{i}~ XZ \otimes Z \otimes X \otimes I_2
\end{array}
\end{eqnarray*}
and 
\begin{eqnarray*}
\begin{array}{ll}
\widetilde{\mathsf{Z}}^4(\vec{f}_1) = X \otimes X \otimes X \otimes X &
\widetilde{\mathsf{Z}}^4(\vec{f}_2) = Z \otimes Z \otimes Z \otimes Z \\
\widetilde{\mathsf{Z}}^4(\vec{f}_3) = X \otimes X \otimes I_2 \otimes I_2 &
\widetilde{\mathsf{Z}}^4(\vec{f}_4) = X \otimes I_2 \otimes X \otimes I_2. 
\end{array}
\end{eqnarray*} 
We choose one of joint eigenspaces $Q(\vec{0})$ spanned by
\begin{eqnarray*}
\left\{
\ket{0000}+\ket{1111},~\ket{0011}+\ket{1100},~\ket{1001}+\ket{0110},~
\ket{0101}+\ket{1010} \right\},
\end{eqnarray*}
and choose $Q_{\min}(\vec{0})$ as 
\begin{eqnarray*}
Q_{\min}(\vec{0}) &=& \left\{
\ket{0000}+\ket{1111}+\ket{0011}+\ket{1100} \right. \\
&& \left. \hspace{10mm} +\ket{1001}+\ket{0110}
+\ket{0101}+\ket{1010} \right\}.
\end{eqnarray*}




\section{Conclusion}

In this paper, we showed a method for enumerating 
all encoding operators in the Clifford group
for a given stabilizer code systematically.
We further classified those encoding operators into
equivalence classes such that EDPs constructed from encoding
operators in the same equivalence class have the same
performance when the input of EDPs is a  probabilistic mixture
of Bell states.
By this classification, we can search EDPs with good performances
efficiently.
As a result, we found the best EDP among
EDPs constructed from $[[4,2]]$ stabilizer codes.
Although in this paper we employed $T=\{ \vec{0} \}$, i.e.,
we abort the protocol if Alice and Bobs' measurement outcomes
disagree, performances of stabilizer EDPs may be
improved by employing $T \neq \{ \vec{0} \}$, i.e.,
we decide whether to abort or perform the error correction according
to the difference of Alice and Bobs' measurement outcome. 
Exploring the potential of $T \neq \{\vec{0} \}$ is
a future research agenda.


\section{Acknowledgment}

This research is in part supported by International
Communication Foundation, Japan.
The authors deeply acknowledge the financial support.


\end{document}